\documentclass{webofc}
\usepackage[varg]{txfonts}   
\usepackage{ptdr-definitions}
\usepackage{hepunits}
\usepackage{heppennames2}
\usepackage{amsmath}
\usepackage{amssymb}
\usepackage{xpatch}
\makeatletter
\xpatchcmd\@HepConStyle
 {\edef\@upcode{\updefault}}
 {\ifdefined\shapedefault\edef\@upcode{\shapedefault}\else\edef\@upcode{\updefault}\fi}
 {}{}
\makeatother
\RequirePackage[colorlinks,citecolor=blue,urlcolor=blue,linkcolor=blue]{hyperref}
\usepackage{adjustbox}
\usepackage{multirow}
\usepackage{lineno}

\newcommand{\MeV}{\ensuremath{\,\text{Me\hspace{-.08em}V}}\xspace}
\newcommand{\TeV}{\ensuremath{\,\text{Te\hspace{-.08em}V}}\xspace}
\newcommand{\GeV}{\ensuremath{\,\text{Ge\hspace{-.08em}V}}\xspace}

\newcommand {\etsumhf}     {\ensuremath{E_{\mathrm{T, sum}}^{\mathrm{HF}}}\xspace}
\newcommand {\cssquare}     {\ensuremath{c^2_{\mathrm{s}}}\xspace}

\newcommand {\nch}     {\ensuremath{N_{\mathrm{ch}}}\xspace}
\newcommand {\nchnorm}     {\ensuremath{N_{\mathrm{ch}}^{\mathrm{norm}}}\xspace}

\newcommand {\ptave}     {\ensuremath{\langle p_\mathrm{T} \rangle}\xspace}
\newcommand {\ptavenorm}     {\ensuremath{\langle p_\mathrm{T} \rangle^{\mathrm{norm}}}\xspace}

\newcommand {\PbPb}  {\ensuremath{\text{PbPb}}\xspace}

\begin{document}
\title{Extracting the speed of sound in the strongly interacting matter created in relativistic nuclear collisions with the CMS experiment}
%
%

\author{\firstname{Cesar} \lastname{Bernardes (on behalf of the CMS Collaboration)}\inst{1,2}\fnsep\thanks{\email{cesar.augusto.bernardes@cern.ch}}
}

\institute{IF-UFRGS, Av. Bento Gonçalves 9500, Porto Alegre, Brasil  
\and
           NCC-UNESP, R. Dr. Bento Teobaldo Ferraz 271, São Paulo, Brasil
          }

\abstract{%
  A hot and dense matter exhibiting collective flow behavior with almost no viscous dissipation has been discovered in ultrarelativistic nuclear collisions. To constrain the fundamental degrees of freedom and equation of state of this matter, these proceedings present an extraction of its speed of sound using head-on lead-lead collision data collected by the CMS experiment at a center-of-mass energy per nucleon pair of 5.02\TeV. The measurement is based on an analysis of the observed charged multiplicity dependence of the average particle transverse momentum in ultra-central events (impact parameter of nearly zero). This variable probes the system temperature as a function of entropy density at a fixed volume. Results are compared with hydrodynamic simulations and lattice quantum chromodynamics (QCD) predictions of the equation of state at high temperatures and small chemical potential. Implications to search for QCD phase transition and the critical point are discussed.
}
\maketitle
\section{Introduction}
\label{sec:intro}

In ultrarelativistic nuclear collisions, many evidences of the production of a state of matter composed of quarks and gluons are observed. This medium has properties compatible with an almost perfect fluid, called quark-gluon plasma (QGP)~\cite{Rhic1,Rhic2,Rhic3,Rhic4}. These collisions are generally described in several stages, with the QGP state as a specific phase modelled by relativistic hydrodynamics~\cite{Heinz}. In these proceedings are presented experimental results~\cite{HIN-23-003} of a new hydrodynamic probe~\cite{Gardim1} in lead-lead (\PbPb) collisions with center-of-mass energy per nucleon pair of $\sqrtsNN=5.02\TeV$, collected by the CMS experiment~\cite{CMS} at the LHC. This innovative technique uses the charged particle multiplicity \nch dependence of mean transverse momentum \ptave at a fixed \sqrtsNN in collisions in which the ions nearly overlap, i.e., collide at a very small impact parameter $b$, dubbed ``ultra-central'' collisions. 

As described in Ref.~\cite{Gardim2}, the speed of sound in the medium is predicted to be related to \ptave and \nch via the relation:
\begin{equation}
\cssquare= \frac{\rd P}{\rd\varepsilon} = \frac{s \rd T}{T \rd s} = \frac{\rd \ptave / \ptave}{\rd \nch / \nch},
\label{eq:speedOfSound}
\end{equation}
where $P$, $\varepsilon$, $s$ and $T$ are the pressure, energy density, entropy density and temperature, respectively. To constrain the equation of state EoS, a simultaneous determination of \cssquare and its corresponding temperature is necessary. It has been shown by hydrodynamic simulations in Refs.~\cite{Gardim1,Gardim2} that $\ptave/3$ is a good estimator of an effective temperature $T_{\text{eff}}$ of the QGP phase. The $T_{\text{eff}}$ can be interpreted as the initial temperature that a uniform fluid at rest would have, possessing the same amount of energy and entropy as the QGP fluid reaches its freeze-out state, when the quarks become bound into particles. Due to the longitudinal expansion and cooling, the $T_{\text{eff}}$ value is generally lower than the initial temperature of the QGP fluid. In these proceedings the \cssquare is extracted for a specific $T_{\text{eff}}$ and the results are compared with theoretical models of relativistic nuclear collisions and predictions from lattice quantum chromodynamics (QCD). 

\section{Measurement method}
\label{sec:method}

The main experimental observable is the \ptave of charged particles in an event as a function of \nch, where \ptave and \nch are measured within the same pseudorapidity ($\abs{\eta}<0.5$) and \pt ranges. Average \pt spectra for $\pt>0.3\GeV$ are measured for events in 50\GeV intervals of the sum of the transverse energy \etsumhf in the CMS forward hadron calorimeter (HF) from 3400\GeV to 5200\GeV, where tracking efficiency and misreconstruction effects are corrected. To avoid any bias in estimating \ptave and \nch, it is necessary to extrapolate the measured \pt spectra to $\pt>0\GeV$~\cite{Gardim1}. The resulting \ptave values from all \etsumhf intervals are then plotted against the corresponding \nch values to form the final observable.


As the extraction of the speed of sound mainly depends on the relative variation of \ptave with respect to \nch (see Eq.~(\ref{eq:speedOfSound})), normalized quantities, $\ptavenorm=\ptave/\ptave^{0\text{--}5\%}$ and $\nchnorm=\nch/\nch^{0\text{--}5\%}$, are used as the primary observables. Here the $\ptave^{0\text{--}5\%}$ and $\nch^{0\text{--}5\%}$ represents the mean transverse momentum and charged multiplicity for the 0--5\% centrality (where centrality is related to the degree of overlap between the two Pb nuclei, with 0\% representing the highest overlap between the two nuclei). Here, the centrality class only needs to be close to that used for the \cssquare determination. Normalizing both \ptave and \nch by their values in the reference event class can minimise most of the systematic uncertainties. 

To extract the speed of sound, the expression that describes \ptavenorm as a function of \nchnorm is taken from Ref.~\cite{Gardim1}, as
\begin{equation}
\label{eq:sound4}
\ptavenorm = \left(\frac{\nchnorm}{\mathcal{P}(\nchnorm)}\right)^{\cssquare},
\end{equation}
where the term $\mathcal{P}(\nchnorm)$ is a function to correct for charged particle multiplicity fluctuations effects for a fixed impact parameter, and its parameters are extracted by fitting a charged particle multiplicity distribution in data~\cite{HIN-23-003}. After extracting the parameters of $\mathcal{P}(\nchnorm)$, a fit to the measured \ptave as a function of \nch in data by Eq.~(\ref{eq:sound4}) is performed to extract \cssquare. 



\section{Results}
\label{sec:results}

\begin{figure}[h!]
    \includegraphics[width=0.49\linewidth]{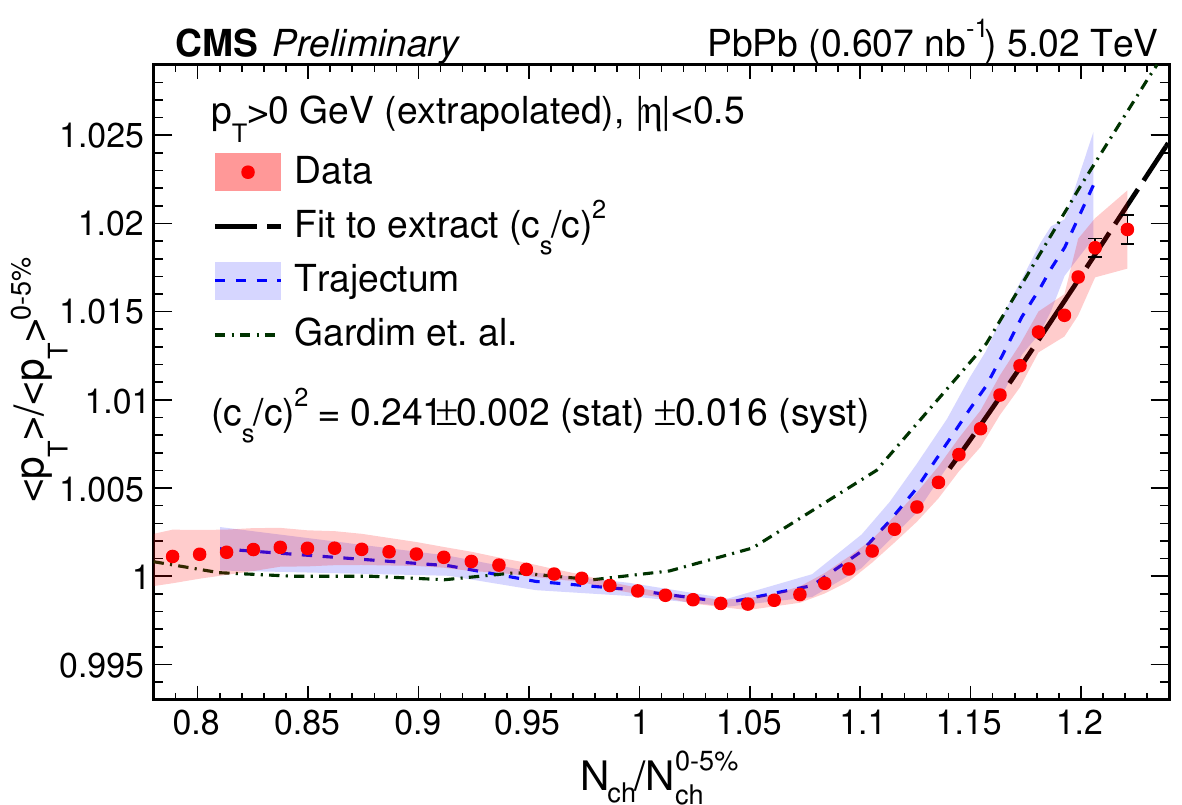}
    \includegraphics[width=0.49\linewidth]{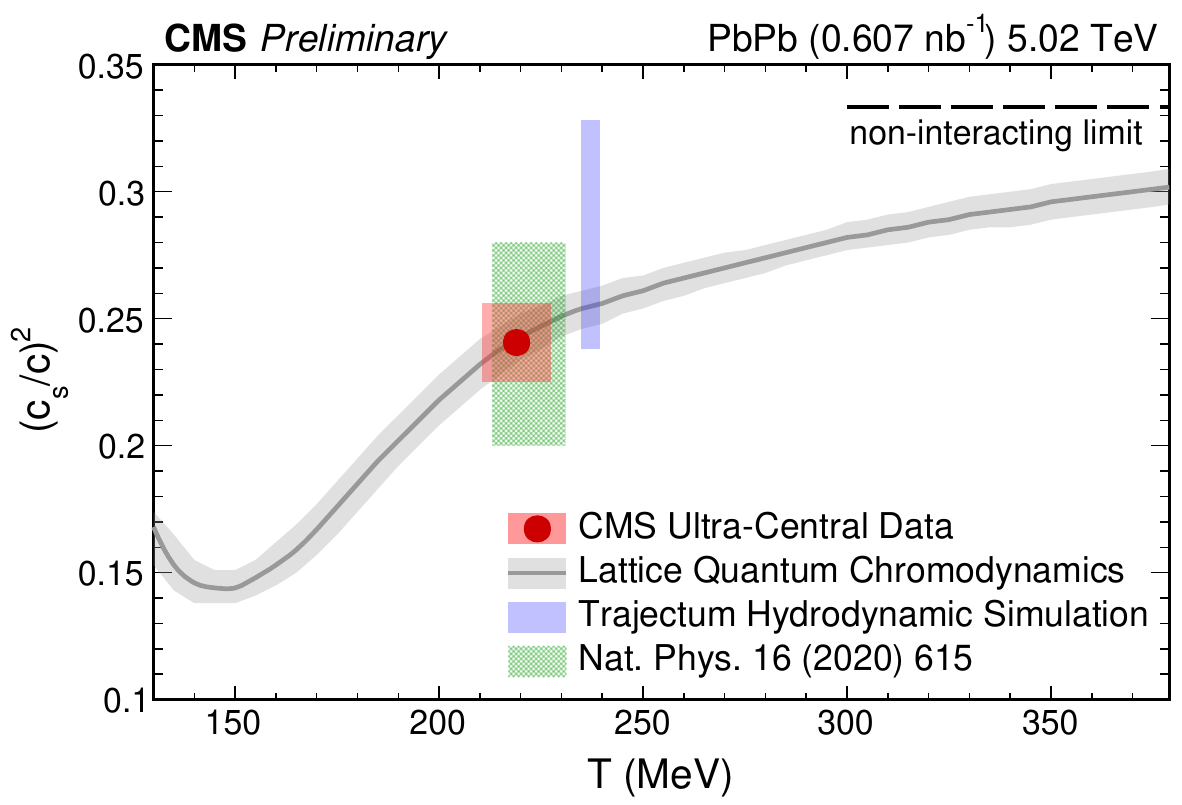}
    \caption{Left: the average transverse momentum \ptave of charged particles as a function of the charged particle multiplcity \nch within the kinematic range of $\abs{\eta}<0.5$ and extrapolated to $\pt>0\GeV$. Both \ptave and \nch are normalized by their values in the 0--5\% centrality class. Bars and bands correspond to statistical and systematic uncertainties, respectively~\cite{HIN-23-003}. Right: The square  of speed of sound as a function of temperature. The size of the red box indicates systematic uncertainties of \cssquare and $T$. To extract the values from Trajectum, the same procedure as applied to data is used. The curve shows the prediction from lattice QCD calculations. The dashed line at the value of 1/3 corresponds to the upper limit for noninteracting, massless gas (``ideal gas'') systems~\cite{HIN-23-003}.}
    \label{fig:fig1}
\end{figure}

The observed multiplicity dependence of the \ptave is presented in Fig.~\ref{fig:fig1} (left). Hydrodynamic simulations from the Trajectum~\cite{Trajectum1,Trajectum2} and Gardim et al.~\cite{Gardim2, Gardim1} models are also shown for comparison. The \ptavenorm value first shows a very weak declining trend toward a local minimum around $\nchnorm\sim1.05$. A steep rise is observed at higher multiplicities, corresponding to ultra-central \PbPb events. The observed trend, including the minimum around $\nchnorm\sim1.05$, is qualitatively consistent with the prediction by the Trajectum model. The model by Gardim et al. also predicts a rise of \ptavenorm at very high multiplicities, with a slope similar to that observed in the data. However, it shows a flat trend at lower multiplicities instead of the local minimum structure around $\nchnorm\sim1.05$ as seen in the data and the Trajectum model. The origin of the observed local minimum is yet to be understood.


To directly extract the speed of sound, the multiplicity dependence of the \ptavenorm data in Fig.~\ref{fig:fig1} (left) is fitted by Eq.~(\ref{eq:sound4}). Because the observed local minimum is not captured by the simplified model in Eq.~(\ref{eq:sound4}), the fit is performed only in the high-multiplicity range of $\nchnorm>1.14$, where the goodness-of-fit value is the best. The final result of the squared speed of sound, \cssquare, is found to be $0.241\pm0.002\stat\pm0.016\syst$ in natural units. The same fit is also performed to the prediction from the Trajectum model, resulting in $\cssquare=0.283\pm0.045$, where the model uncertainty is determined within the allowed parameter space constrained by a global Bayesian analysis~\cite{Trajectum1,Trajectum2}. In recent studies~\cite{Trajectum3}, it was shown that the slope of the steep rise of \ptave vs \nch can depend, in addition to the EoS, on the simulation parameters used to describe the initial transverse energy density and also the centrality selection definition. Regarding the former, it is important to note that correlations between these parameters and \nch were not taken into account in Ref.~\cite{Trajectum3}, which can exclude the existence of such large parameters and consequently do not result in considerable influence in the slope.  For the latter, based on studies using the CMS data, in order to have stable results, it is recommended to use a large $\eta$-gap between the definition of centrality and the measured  \ptave and \nch, as performed in the present analysis.

The $T_{\text{eff}}$ is measured as $\ptave^{0\text{--}5\%}/3 = 219\pm8\syst\MeV$, where the statistical uncertainty is negligible. Figure~\ref{fig:fig1} (right) shows \cssquare as a function of $T$, with the CMS data point at $T_{\text{eff}}$. The results are compared to lattice QCD predictions~\cite{Lqcd}, the Trajectum model, and the \cssquare value extracted in Ref.~\cite{Gardim2}. The new CMS data allow for an unprecedented level of precision in the experimental determination of the speed of sound in this effective temperature. The results exhibit excellent agreement with the lattice QCD prediction, with comparable uncertainties. Thus, these findings provide compelling and direct evidence for the formation of a deconfined QCD phase at LHC energies. This relatively simple procedure can be applied for different colliding systems and energies and, if systematically compared with calculations from lattice QCD, can be used to help in the search for the QCD phase transition and the critical point. Is important to note that first attempts to extract the speed of sound in ultrarelativistic collisions at lower center of mass energies were performed by analyzing rapidity distributions~\cite{cssquare1, cssquare2}, extracting values for \cssquare of similar order as in the present work.


\section{Conclusions}

These proceedings present a measurement of a new hydrodynamic probe resulting in the most precise extraction to date of the speed of sound of $0.241\pm 0.002\stat\pm0.016\syst$ (in natural units) in ultrarelativistic nuclear collisions at an effective temperature of $219\pm8\syst\MeV$. This is performed by determining the dependence of the mean transverse momentum on the total charged particle multiplicity in nearly head-on lead-lead collisions with center-of-mass energy per nucleon pair of $\sqrtsNN=5.02\TeV$. The excellent agreement of lattice quantum chromodynamics (QCD) predictions with the experimental results provides strong evidence for the existence of a deconfined phase of quantum chromodynamics matter at extremely high temperatures and small chemical potential. This method can contribute to the search for the phase transition and the critical point in high-density QCD.

\section*{Acknowledgements}

This material is based upon work supported by FAPESP under Grant No. 2018/01398-1, by FAPERGS Grant No. 22/2551-0000595-0, and by CNPq Grant No. 407174/2021-4.

%

\begin{thebibliography}{}
%
%

\bibitem{Rhic1} 
STAR Collaboration, Nucl. Phys. A \textbf{757}, 102 (2005)

\bibitem{Rhic2} 
PHENIX Collaboration, Nucl. Phys. A \textbf{757}, 184 (2005)

\bibitem{Rhic3} 
BRAHMS Collaboration, Nucl. Phys. A \textbf{757}, 1 (2005)

\bibitem{Rhic4} 
PHOBOS Collaboration, Nucl. Phys. A \textbf{757}, 28 (2005)

\bibitem{Heinz}
U. Heinz, R. Snellings, Ann. Rev. Nucl. Part. Sci. \textbf{63}, 123 (2013)

\bibitem{HIN-23-003}
CMS Collaboration, CMS Physics Analysis Summary, HIN-23-003, \url{https://cds.cern.ch/record/2870141} (2023)

\bibitem{Gardim1}
F. G. Gardim, G. Giacalone, and J. -Y. Ollitrault, Phys. Lett. B \textbf{809}, 135749 (2020)

\bibitem{CMS}
CMS Collaboration, JINST \textbf{3}, S08004 (2008)

\bibitem{Gardim2}
F. G. Gardim, G. Giacalone, M. Luzum, and J. -Y. Ollitrault, Nature Phys. \textbf{16}, 615 (2020)

\bibitem{Trajectum1}
G. Nijs and W. van der Schee, Phys. Rev. C \textbf{106}, 044903 (2022)

\bibitem{Trajectum2}
G. Giacalone, G. Nijs, and W. van der Schee, arXiv:2305.00015 (to be published)

\bibitem{Trajectum3} 
G. Nijs, and W. van der Schee, arXiv:2312.04623 (to be published)

\bibitem{Lqcd}
HotQCD Collaboration, Phys. Rev. D \textbf{90}, 094503 (2014)

\bibitem{cssquare1}
B. Mohanty, and J. Alam, Phys. Rev. C \textbf{68}, 064903 (2003)

\bibitem{cssquare2}
NA57 Collaboration, J. Phys. G: Nucl. Part. Phys. \text{31}, 1345 (2005)

\end{thebibliography}
%
%

\end{document}